# Prediction Accuracy and Autonomy

Anton Angwald, Kalle Areskoug, Alan Said

*University of Gothenburg, Sweden*
*antonangwald@icloud.com, areskoug.kalle@gmail.com, alansaid@acm.org*

**Abstract**
The tech industry has been criticised for designing applications that undermine individuals' autonomy. Recommender systems, in particular, have been identified as a suspected culprit that might exercise unwanted control over peoples' lives. In this article we try to assess the objectives of recommender system research and offer a nuanced discussion of how these objectives can align with users' goals. This discussion employs a qualitative literature survey connecting the dots between relevant research within the fields of psychology, design ethics, interaction design and recommender systems. Finally, we focus on the specific use-case of YouTube's recommender system and propose design changes that will better align with individuals' autonomy. Based on our analysis we offer directions for future research that will help secure rights to digital autonomy in the attention economy.
**Keywords**
Recommender systems, autonomy, design ethics, user studies, evaluation metrics

## 1. Introduction

Recommender systems in the entertainment domain have evolved to focus on maximising some engagement metric, e.g., *watch time*. Even though it seems reasonable to assume, that if a particular user has watched one hour worth of music videos, this user values music, researchers have noted several issues with this approach [1]–[7]. The central problem being, what in psychology is referred to as an intention-behaviour gap [8]. Simply put, people do not always do what they intend to do. Behavioural user data such as watch time are also subject to feedback loops in recommender systems. Feedback loops are caused by the fact that the data being used to make recommendations is influenced by the recommendations themselves. This confounds the users' intended behaviour with behaviour that might have been shaped by the system's persuasive ability [9]–[11]. A user who ended up watched one hour of music videos might have originally wanted to do something else. This user, when finally quitting the application, did perhaps do so with regret, feeling they wasted their time. Studies have shown it is common for users to complain that social media services waste their time and that they often regret using them [12]–[15].

In this work, we explore the question of *how design of recommendation systems for entertainment services can align with users' autonomy*. We specifically focus on psychological theories of agency, user studies on entertainment recommender systems and conceptualisations of user-centered value. Reviewing literature on these subjects we try to form a coherent picture of problems that recommender system designers within the entertainment domain need to solve in order to secure individual users' right to autonomy.

The practical importance of this work is further stressed by public concerns with social media addiction [16]–[20]. However, this criticism, along with other issues of user autonomy, are not solely linked to recommender systems. There are other influencing factors outside of recommender systems, in other types of interaction design [21] and especially in relation to the social psychological aspects of the network [22], [23].

To reduce confounding social psychological factors, while still relating to a highly influential media platform, we will, when appropriate, focus on the specific case of YouTube's recommender algorithm since the YouTube application is content-based rather than community-based. Even though the ideas outlined in our work can be extended to analyses of other recommender systems, focusing on YouTube has a value on its own since the platform has more than 2 billion monthly logged-in users, from more than 100 different countries. Google's CPO estimated that YouTube's recommendations drive 70% of the watch time on the platform [24].

## 2. Related work

Autonomy & Personal identity has been identified as one of six key areas of ethical concern in research on recommender systems [25]. Varshney [26] suggested that recommender systems may undermine sense of agency by relying too much on behavioural data in measuring the effectiveness of the system. The work argues that it is essential to also include the notion of autonomy when evaluating recommender systems. Ekstrand and Willemsen [5] discuss the reliance on behavioural data in recommending content to users. An advantage of implicit data is that it is more available, as it consists of automatically collected data such as clicks and watching time. Another advantage is that it better predicts future behaviour in comparison to explicit ratings from users. According to Ekstrand and Willemsen the discrepancy between what users say (explicit data) and what users do (implicit data) can be explained either by (a) the user does not understand their true desires, or, (b) the user is dissatisfied with their behaviour and wishes to change it. These two options provide an introduction to how the concept of user autonomy is intimately linked to user value. If (b) is true, optimising for users' behaviour instead of stated preferences can be reasonably argued to undermine autonomy, since it acts against the users' own goals and wishes. However, if (a) is true, the objective for recommendation systems to retain individual autonomy becomes philosophically problematic. Should recommender systems aim to understand the "true" desires of users and optimise for these? Or is this a form of paternalistic stance that undermines personal autonomy by acting on the assumption that the individual is incapable of understanding their own best interests and therefore of taking their own decisions?

James Williams, dedicating a book to the topic of information technology and autonomy [27], compares recommender systems to a GPS-system whose goal should be to guide us through digital space rather than through physical space. Entering an address on a GPS-system and ending up somewhere else would be evident of a faulty GPS-system and we should treat recommender systems by the same standard. Knijnenburg et al. [3] argued that the primary goal of recommender systems should be to increase user experience and that this is not the same thing as maximising prediction accuracy. They developed a framework for measuring user experience and evaluated the connection between user experience and prediction accuracy. Their results give several examples of how there is a poor relationship between user experience and prediction accuracy. While prediction accuracy is measured implicitly, through for example clicks and interaction times, they measure user experience explicitly, through user testing of systems accompanied with interviews and surveys. Several other researchers have also challenged the assumption that algorithms which better predict behaviour lead to better recommender systems [2], [4], [6], [28].

## 3. Cognitive Science Perspectives
## 3.1. Psychology of User Autonomy

In this section we will give perspectives on recommendation systems from cognitive science, more specifically we will look at psychology of autonomy, exploring studies on three subcategories: sense of agency, self-regulation and habitual behaviour. When appropriate we will relate this research to user studies on entertainment recommender systems.

### 3.1.1. Sense of Agency

Autonomy is one of the main areas of ethical concerns in recommender systems. One psychological research article gives the following definition: "Autonomy refers to self-government and responsible control for one's life." [29]. One way of approaching autonomy without having to indulge in metaphysical debates on free will is to, instead, talk of a personal perception of autonomy; sense of agency. Sense of agency can be further divided into feelings of agency and judgement of agency [30]. Feelings of agency is in-the-moment perception of agency and is linked to low-level sensorimotor processes. As an example, feeling in control when using an application, being able to click a button and seeing the interface react. Judgement of agency is a post-hoc perception of agency, estimating how much one was in personal control in a previous situation. It is linked to higher-level cognitive processes, integrating contextual information as well as background beliefs [31]. Both of these refer to the *subjective notion* of being in control rather than *actually* being in control. Sense of agency is therefore not the same as actual agency. They are, however, related. For example, when interacting with technology, higher levels of automation lead to a decreased sense of agency [30]. Klobas et al., [32] conducted an interview study with participants that had self-reported a problematic bond with YouTube. One recurring denominator reported by almost all of the participants as a problem with YouTube, was in situations where the sense of agency decreased due to automation.

### 3.1.2. Self-regulation

One example where a user might have had high feelings of agency but low judgement of agency is when reflecting back, wondering why they just spent two hours on watching YouTube videos when they had the goal of spending just ten minutes. Using applications in ways that are later regretted is a problem that has been widely reported by users [12]–[15]. Research on digital wellbeing has used the concept of "lagging resistance" to describe the self-reported tendency of users wanting to quit using an application but not wanting to do so just yet [15], [33]. Lagging resistance can therefore be described as a conflict between instant gratification and long-term goal-achievement. To overcome the problem of lagging resistance the user needs to inhibit desire in order to perform a goal-directed action. Such an inhibition of desire demands self-control. In accordance with previous literature we will use the concept of self-control to refer to the act of inhibiting desire while using the more inclusive concept of self-regulation to denote regulation of behaviour according to one's own goals [34].
Hofmann et al. [34] study how individuals with various levels of self-control differ in how they manage to prevent themselves from acting on desires in conflict with their goals. Their study suggests that individuals with high levels of self-control are not better at resisting desires, rather that they manage to avoid situations where conflicting desires appear in the first place. Other studies support the thesis that individuals with lower inhibitory abilities as well as lower attentional control (such as people with ADHD/ADD) are worse at practicing self-control which also means that they are at higher risk of suffering from what some researchers call social-networks-use disorder. Wegmann et al. [35] define "social-networks-use disorders" as habitual usage triggered by impulsive responses to cues. The authors write: "A dominance of the impulsive system is assumed to induce approach tendencies towards potentially gratifying options while neglecting long-term risks, which may result in risky behaviour such as drug consumption". Their study suggest that the same approach tendencies can be a driving factor to using applications in a way that is later regretted.

Repeatedly failing in resisting desire can result in a state of learned helplessness, referring to an acquired belief that one is unable to change their situation. This can lead to a negative spiral [36] since people in a state of learned helplessness, actually decrease their efforts to resist desire. Similar to the notion of learned helplessness, studies have shown that addicts experience a spiralling failure in self-regulation after what Marlatt called an "abstinence violation effect" [37], [38]. This refers to how minor violations of self-regulative behaviour might lead to a total collapse of self-regulation. Other studies have shown that repeated situations in which self-regulation is required, depletes cognitive resources needed to exhibit self-regulation. Repeated exposures of tempting stimuli are thus likely to lead to failure in self-regulation. This concept is also referred to as ego-depletion [37]–[39]. This is in accordance with the results from [34] which suggest that individuals successful in self-control succeed by avoiding tempting situations in the first place rather than by having greater abilities to inhibit impulsive behaviour. The same depletion of cognitive resources relevant for self-regulation has also been shown to be induced by information overload [40].

### 3.1.3. Habitual Behaviour

Gollwitzer & Sheeran [8] found in their meta-analysis of meta-analyses that intentions only explained 28% of the variance in behaviour. They called this an intention-behaviour gap. Even though the exact size of the gap was difficult to estimate, the intention-behaviour gap was by the smallest estimates still large enough to show that people frequently behave against their own intentions. The authors suggested that the intention-behaviour gap might be due to habitual behaviour. In a dual processing framework, habitual behaviour can largely be described as fast, automatic and unconscious while non-habitual behaviour is slow, deliberate and reflective [39]. In similarity to impulsive behaviour, habitual behaviour can function as a sequence of actions which are enacted as an automatic response to external stimuli [39]. This is why Schnauber-Stockmann et al. [41] say that habits function within the impulsive system, which can be contrasted to the reflective system. While processing within the reflective system has the disadvantage of relying on potentially effortful deliberation, it has the advantage of being oriented towards long-term goals and abstract values. The impulsive system is oriented towards immediate gratification and relies on cognitive heuristics which can be described as quick-and-dirty strategies for making decisions, highly susceptible to biases [41].

If recommender systems promote habitual usage, they in turn reduce individuals' opportunities for reflection. These opportunities are crucial for turning attention to our own mental activities in order to "call our beliefs and motives into question." [27], [42]. Patterns of habitual usage might therefore also undermine the ability to form a personal identity, which is connected to autonomy [25].

### 3.1.4. User Case Studies

The reflective system is characterized by high levels of self-regulation while the impulsive system is, as previously discussed, related to technology usage that is later regretted. With this in mind, we can expect that negative media behaviour would largely be habitual behaviour. This is exactly what a user study on smartphone behaviour [15] suggested. They found that people sometimes experienced a loss of autonomy when they were using their smartphones and in these cases, participants highlighted the habitual and automatic nature of their usage. Moreover, they found that "Lack of control was rarely attributed to active failure to resist in-the-moment, but rather to unconscious habit." A similar finding comes from Van Deursen et al. [43]. They found that addictive smartphone behaviour was often associated with habitual smartphone use. In another design case study, 120 heavy users of YouTube participated [44]. The authors evaluated how YouTube's recommender algorithm affects users' sense of agency, as well as asking users what they thought of design proposals. The study found that irrelevant recommendations decreased users' sense of control but additionally, that for roughly half of the participants even relevant recommendations could decrease their sense of control. Relevant recommendations decreased control in situations when the user was using the app habitually or at an unsuitable time (often late at night). The authors explained this issue as the result of recommendation algorithms being good at solving a local optimisation problem ("what should I watch on YouTube?") while failing to solve a global optimisation problem ("should I watch YouTube or not?"). In relation to YouTube's recommendation system, Klobas et al. [32] found that the behaviour of clicking on related videos (that are chosen by recommendation algorithms) was strongly correlated with compulsive YouTube use. Even if sessions sometimes begin with users watching a goal-directed, productive video it becomes irrelevant to the original intention as the user clicks related recommended videos, each one deviating more from the starting video [32]. In Lukoff's study [15] on smartphone behaviour, users' intention was also gradually diminished, which could be explained by design features promoting habitual usage.

### 3.1.5. Summary

According to the psychological literature, if recommendation algorithms that are evaluated on measuring behaviour (implicit data) rather than intent (explicit data) have higher prediction accuracy it is not surprising. The discrepancy between explicit and implicit data could merely reflect the already demonstrated intention-behaviour gap. If this gap is due to habitual behaviour, as the psychological literature suggests, and we are to trust in the dual processing theories of decision making, the gap reflects inherent cognitive weaknesses in decision making. If the ultimate goals of recommender systems are to help users with decision making, relying on behavioural data might therefore be contra productive. These systems should help users bridge the intention-behaviour gap rather than achieving a high prediction accuracy by taking advantage of (or even increasing) the intention-behaviour gap. However, note that this argument presupposes that goal-directed decision making is more valuable than impulsive decision making. The dual and sometimes conflicting nature of decision making makes the notion of autonomy difficult to interpret. Perhaps one way to resolve this is to consider what type of autonomy that users deem valuable. This is what we will explore in the next section.

## 3.2. User Autonomy & User Value

In this section we will briefly discuss problems of correctly determining what type of content is valuable to users. We will then summarise various user-centred design proposals that aim to combat problematic technology behaviour.

### 3.2.1. What Users Really Want

Williams [27] states "Whether irresistible or not, if our technologies are not on our side, then they have no place in our lives." but what "on our side" actually means is harder to define. This taps into an active ethical debate on persuasive technology and nudging [45]–[47]. General positions in this debate are characterised by Lyngs et al. [45] in their "...fictive dialogue between senior executives at a tech company aimed at helping people live the life they 'really' want to live". Even if this fictive dialogue is quite an unorthodox way of approaching this ethical question, we find it to be of great pedagogical value for introducing the various positions in this debate for the unfamiliar reader. Below follows an excerpt from the paper [45]:

> *"But what does any of that actually mean? How can we be sure that we are giving users what they really want? What we need, my friends, is a clear answer to this question; a new metric towards which all our services should be geared; a new optimisation metric for life. So come on, hit me with your ideas!*
> ***Randy:*** *I'm going to stop you right there, sir, if I may. What's wrong with our existing systems? We infer what users want from what they do and what other people like them do. If they spend every spare second watching cat videos, then our algorithms should give them more cat videos. If they keep watching them, that means our algorithms got it right. If they don't like them they will stop looking at them. Our algorithms will then show them less in the future ...*
> ***Harald:*** *Woah there. I totally disagree. People are slaves to simple reward functions inherited from our evolutionary past. We know how to hack these reward systems, so if we leave people to their own devices (no pun intended) they will simply do whatever our algorithms nudge them to do. That might be binge-watching cat videos and ordering takeout pizza. It probably won't be filling in their tax returns or exercise ...*

*Nichola: But we could be nudging them to do those things instead! Even better, we could nudge them to do something truly worthwhile, like reading poetry, or contributing to science, or meditating on the miracle of their very existence!"*

### 3.2.2. Design Proposals for Helping Users

We could base user value on assumptions of what is meaningful human behaviour. Although, this approach might be misguided. The study by Lukoff et al. [15] showed that even when users engage in productive or goal-directed behaviour, they experience it as meaningless and unsatisfactory, if it is habitual behaviour. Persuasive interfaces aimed to nudge users to fulfil long-term goals might thus fail to increase user value if users are pushed to engage in these activities without reflection. Perhaps then, recommender systems need to decrease habitual usage and nudge the user towards more reflective behaviour. Hiniker et al. [13] and Shin & Dey [48] show methods to detect when a user might be interacting habitually rather than intentionally. Inhibiting extensive app usage in those situations might better optimise user value. Another potential approach is proposed by Cheng et al. [49]. They construct a model that from two minutes of behavioural data can predict users' intention of the session, perhaps this model could be used to prevent erosion of intention during a usage session.

Another way to avoid the paternalistic issues of nudging is to empower users to modify designs according to their own preferences. In Lukoff's study of the YouTube platform [44], user participants expressed that available customisation settings helped in combating problematic usage, but that they wanted more ability to customise the recommendations and the interface. The authors' main suggestion is to include a customisable interface with various degrees of control. For example, enabling users to switch between a Focus mode and an Explore mode.

Another proposed solution is an intervention mechanism that will force the user to reflect over his/her usage. The user would be required to solve a cognitive task such as a puzzle in order to continue using the application. This might lead to combating the problem of habitual usage, as users are forced to become more aware of their usage [50]. Other research has proposed similar external mechanisms, such as enabling users to set self-imposed limits on what time of day or amount of time that they can access an application [13].

Other proposed solutions concern changing recommender systems in order to avoid the need for a solution in the first place. Ekstrand & Willemsen [5] argue that explicit ratings (users' self-expressed desires) should continue to be included in the recommendation process alongside implicit data. In accordance with this, the study by Lukoff et al. [15] shows that utilising explicit user ratings is a valid approach to measuring meaningfulness. Perhaps the easiest way to avoid making assumptions about what users really want is by doing just this, asking the users themselves. At least in this case, the assumptions are the users' own assumptions. Therefore, even if it might not optimise user value it should at least optimise autonomy. One way in which this could be done is suggested in a paper from Twitter [51]. The primary aim of this work is to directly respond to the issues outlined by Ekstrand & Willemsen [5] by developing a more correct operationalisation of user value. Milli et al. [51] combines different types of data by weighing them differently. Based on the assumption that explicit data better corresponds to user value, they give higher weights to types of data that are more explicit. For example, if the user clicks to view a tweet, this data has a low weight but if the user clicks the button "See less often" this data is given a high weight.

## 3.3. The Current YouTube Recommender System

In 2019, a Google paper on the YouTube recommendation system [52] proposed and tested a Multi-gate Mixture-of-Experts (MMoE) system architecture. This system would take both engagement objectives and user satisfaction objectives into account. We propose, in line with previous discussion, that these are objectives in the interests of different parties, engagement objectives being of primary interest of the service provider and user satisfaction objectives of primary interest to the user. The paper therefore suggests that YouTube's recommender system utilises both explicit and implicit data. However, the specific balance between these two different objectives is not given, and it seems like this, to a certain degree, is up to the service provider. The degree seems to be limited by a lack of data related to user satisfaction (i.e., ratings and survey responses).

Detailed information about YouTube's recommender system is not available, for understandable reasons. The system represented in the Google paper likely does not represent the current recommender system in use as YouTube continuously develops their recommender system [53]. Therefore, it might be the case that changes that we propose to YouTube's recommender system are already in place, or non-applicable.

## 4. Discussion

Maximising user utility is easier said than done. First of all, there exists the problem of finding the right technique, utilising the right type of data and the right type of evaluation metric. Maybe more pressingly, we need a definition of user utility, and this leads us to having to define what is the "right thing to do" for any particular user, which traps us in the fictive dialogue of Lyngs et al. [45]. Even if this question needs to be addressed eventually, we argue that the research previously outlined in this paper supports the stance that recommender systems should avoid promoting habitual behaviour [8], [15], [41]–[44], [25], [26], [34]–[39] and that there are sound reasons to believe that explicitly stated user preferences correspond better to autonomy concerns [1]–[3], [5]–[8], [26], [27], [51], in comparison to behavioural data.

### 4.1. User-centric Design for Recommender Systems

The solutions proposed in Google's paper [52] and in the paper by Lukoff et al. [44] as well as the experiments on Twitter show a promising step in the right direction. We will now critically assess these solutions by taking advantage of the theoretical background we have outlined in the earlier part of this paper.

Based on previous research [1]–[3], [5]–[8], [26], [27], [51] we have shown that recommendations relying on implicit data can decrease user autonomy as compared to more explicit data. In light of this, it would be reasonable for YouTube to implement a similar approach to the one outlined in Milli et al. [51]. In this system, data that is more explicit, such as user rating, is being valued higher than less explicit data, such as viewing time. In YouTube's case, users' engagement data on a video that is retrieved by actively searching for it can be legitimately assumed to better represent users' intentions, as compared to data on a recommended video [15], [32], [39], [41], [43], [44], [50]. Because of this, engagement data on a video retrieved from a user search query should be valued higher than engagement data on a video retrieved from a recommendation. This might already be the case, but due to a lack of transparency we cannot know.

YouTube should also extend their possibilities for explicit user feedback. This would decrease the problem of data sparsity for assessing user satisfaction. In the current video interface, there are two major options for explicit feedback, thumbs-up and thumbs-down. This user feedback impacts recommendations [10], [52], but due to the ambiguous nature of these buttons, there might exist a discrepancy between what the action actually means and what the designer expects it to mean [51]. A simple solution to strengthening the validity of this explicit feedback is to replace these buttons with one for "Show more often" and one for "Show less often", the explicit instructions on what the button actually does can be expected to decrease the gap between what the action means and what the designer expects it to mean. This would also increase transparency by making it clearer for the user that the action impacts what videos are recommended. As supported by YouTube users' expressed desire for higher customisability [44], readily available buttons that clearly communicate an opportunity for user customisability might be used more often. Because of this, these buttons would not only provide more reliable data, they could also provide larger quantities of data, which would reduce problems of data sparsity.

## 4.2. User Customisability

The two main problems that we have identified in relation to user autonomy can be categorised as:
1. Excessive usage
2. Unsatisfactory usage

While unsatisfactory usage is more linked to local optimisation objectives, the two problems are entwined as a user might categorise something as excessive usage because it is unsatisfactory. They might also categorise something as unsatisfactory because it feels excessive. Since it is difficult to have an objective definition of what is satisfactory and what is not [45], these problems should be addressed from the perspective of user studies as well as discussions specifically set in the context of technology usage. The proposals outlined in the preceding section have mainly addressed unsatisfactory usage (2). If opportunities for explicit user feedback are only in-app, their satisfaction objective only covers local goals. Qualitative user studies like the ones we have previously surveyed [15], [44] better address users' global optimisation problem.

When it comes to excessive usage, the psychological concepts we have previously discussed are highly relevant and designers should mitigate the risks for self-regulation failure in accordance with psychological literature. However, this might involve evaluating one type of user value over another. The concept of "lagging resistance" [33] relates to the conflict between instant gratification and long-term goal achievement. A possible explanation to how this conflict might function that is consistent with the notion of learned helplessness [36] and the abstinence violation effect [37], [38] is by a dissonance between feelings of agency and judgement of agency. Strong feeling of agency might be related to not being prevented from making an instantly gratifying choice and strong judgement of agency might be related to having successfully avoided instantly gratifying options in order to pursue long-term goal-directed behaviour. Therefore, designers might have to make the choice of optimising for feelings of agency or judgement of agency. Some design proposals by Lukoff et al. [44] might reduce feelings of agency while improving judgement of agency, and this is also an issue of several possible design proposals such as lock-out mechanisms. This brings us to the ethical question of which type of user agency that should be valued most. Psychological literature suggests that feelings of agency can be tied to addictive behaviour but also that it has strong ties to satisfaction. When possible, solutions that do not directly decrease feelings of agency therefore have the advantage of avoiding this trade off. Another problem with this trade-off is that users are different. Several participants in the study by Lukoff et al. [44] state that they use YouTube for different purposes at different times, sometimes wanting to be entertained and sometimes wanting to be able to focus. The psychological literature on social media addiction also suggests different needs for different types of people with regards to design. An addictive design feature for one person might simply be fun for another person. Because of this, Lukoff proposes high customisability for the user, one proposal being a discover mode and a focus mode in which the focus mode offers less or no recommendations. While we agree that high customisability is good, we should think of the nature of that customisability. We propose that increased feelings of agency might lead to higher feelings of responsibility. If a user has higher feelings of agency and fails to exercise that agency, it leads to a higher self-attribution of that failure which in turn lowers the hindsight judgement of agency. This leads to the apparent contradiction that increased feelings of agency can in certain situations reduce the belief of self-control abilities. Moreover, a reduction of this belief can lead to actual self-regulation failure [36]–[38].

External lock-out mechanisms are sensitive to this problem. An easily bypassed external lock-out mechanism that, for example, can be unlocked with a password, might both increase the users' feelings of agency and help the user exercise self-control. However, if the mechanism is easily bypassed, it is likely that the user will start to habitually ignore the lock-out mechanism and without reflecting, enter the password. The lock-out mechanism will then actually make it worse for the user because not only does it fail in helping the user to take a break, it also makes the user feel guilty for not taking a break. In the best scenario, this will lead to temporary ego-depletion. In the worst scenario, this behaviour over time might lead to learned helplessness and therefore reduce the users' self-regulation ability.

Because of issues with giving users high customisability, we propose the following considerations. First of all, it is essential that the user should not be given more choices if the choices are unlikely to actually make a difference to the user. If external lock-out mechanisms do not actually make significant reductions in compulsive behaviour they are more likely to do harm than good. Secondly, when customisability can make reductions in compulsive behaviour, we propose that they can be framed in a way to increase user sense of agency while decreasing the likelihood of experiencing guilt. We think that the fact that sense of agency does not perfectly correspond to actual agency can be utilised. One example of this is default bias. Having the default option being the current YouTube layout but having a continuously present option saying "enter focus mode", rather than having to choose between focus and entertain mode, gives different results for users. Only having the option to actively increase "focus" empowers users to make this choice while not introducing a sense of guilt for users who do not make this choice. An added benefit of such a user customisability is that YouTube gets one more point of explicit data. Knowing when people switch from the default mode to the focus mode is useful in understanding when people feel distracted by the recommendations. Analysing this data might give a better understanding of when users gain utility from recommendations and when they do not.

One external solution that is likely to bypass the problems mentioned in the two preceding sections is the one proposed by Park et al. [54]. They propose giving the user more autonomy by forcing them to reflect on their usage by inhibiting habitual behaviour through a cognitive task. We think this is likely to work but it is also likely to be annoying. This problem could perhaps be solved by optimising the difficulty and varying the type of task. However, while this option can be an empowering tool for users, it does not help in aligning technology to users' goals in the first place. Aside from the fact that Williams [27] makes a valid philosophical point that this should be a necessary ethical requirement for technology in the first place, it is also a temporary solution. If users need to employ empowering intervention mechanisms that deal with the problems of aversive technology, they need to be in a constant state of learning and discovering in order to be able to adapt to an ever-changing technological landscape.

## 5. Conclusion

We have highlighted how certain aspects of entertainment recommender systems can cause a problem for the individual autonomy of users. The primary problem we have discussed is how recommender systems try to predict the intentions of users from their behavioural data rather than from their expressed desires. Through assessing psychological literature on autonomy and user studies on entertainment services we have shown how users' behaviour is an inaccurate reflection of their intentions.

With this in mind, we have explored solutions to the following question: *"How can the design of recommendation systems for entertainment services align with the individual right to autonomy?"*

These solutions have been both of preventive and corrective nature. The corrective solutions have been focused on offering users' more customisability. The preventive solutions have been focused on gathering more data that correspond better to users' intentions. We have also shown how higher customisability can provide user data that can be expected to correspond relatively well to users' intention.

Answering the question stated above will be a gradual undertaking and we have shown promising starting points for this venture. We have suggested that it is essential that users' right to autonomy is discussed in relation to users' values. This is to ensure that good-intentioned solutions aimed to increase users' autonomy do not result in unsatisfactory experiences.